# Generalized Action-based Ball Recovery Model using 360° data

Paper Track

Ricardo Furbino Marques do Nascimento[1] and Hugo Rios-Neto[2]

## Introduction

Even though having more possession does not necessarily lead to winning, teams like Manchester City, Liverpool, and Leeds United notably have tried to recover the ball quickly after they lost it over the past few years. Nowadays, some of the top managers in the world apply high-pressing styles, and concepts such as the five-second rule, usually credited to Guardiola, have been spreading out [9][10], becoming a fundamental part of how lots of teams have played over the recent years. Expressions like "don't let them breathe" and "get the ball back as soon as possible" are often heard in the media [4][5][6], but what are the actions that most lead to a change in possession? What is the influence of a team's positioning on the ball recovery? Which are the players that more often collapse when under pressure? Can we evaluate the defensive dynamics of teams that do not necessarily press the player in possession as intensely as those mentioned above? We try to answer those and other questions in this paper by creating a Generalized Action based Ball Recovery model (GABR) using Statsbomb 360° data.

Recent work in the field has focused on evaluating pressing dynamics and their effects on recovering the ball. Robberechts' Valuing Pressure decisions by Estimating Probabilities (VPEP) framework [2] was the first to quantify the effectiveness of pressing in different phases of the game and its risk-reward trade-offs. It used event data, more specifically Statsbomb's, as it was the first provider of distinct pressure events. [3] extended the work in [2] to tracking data, building a framework that could both identify pressing scenarios and later evaluate them, using Pitch Control features to better capture some of the spatial dynamics and increase the assertiveness of the model. Both took inspiration from Decroos et al.'s Valuing Actions by Estimating Probabilities framework [1], a state-of-the-art action-based evaluation framework that estimates an action's contribution to short-term scoring and conceding probabilities.

Every team's objective is to eventually recover the ball when out of possession. A team's ball recovery strategy may vary based on their own quality, their opponent's, the score, and where they plan to recover the ball, among other reasons. This means a team may try to recover the ball as quickly as possible, pressuring the player with the ball, or they may try to recover it afterward if they believe it can yield them better opportunities when again in possession. However, even when not directly pressing the player with the ball, a certain strategy is being deployed to recover it. Also, as noted in [11], even though a team might appear to be "parking the bus", its players are ready to go forward as soon as they recover the ball.

[1] Undergraduate Computer Science Student at Universidade Federal de Minas Gerais (UFMG). Belo Horizonte, Brazil. ricardofurbino@dcc.ufmg.br.
[2] Data Scientist at Clube Atlético Mineiro and Computer Science Master's Student at Universidade Federal de Minas Gerais (UFMG). Belo Horizonte, Brazil. hugoriosneto@dcc.ufmg.br.

The main objective of our work is to assess the off-ball dynamics of teams out of possession by estimating its effects on the probability of a ball recovery. Consequently, we do not need to restrict ourselves to moments when the defending team is exerting pressure on the player with the ball. Furthermore, we can also estimate the resistance of on-ball players to these dynamics. Our work is heavily inspired by [2] and [3], serving as a natural extension and generalization from [2] to Statsbomb's contextual event data, called 360º.

## Framework

Research in Football Analytics has tried to solve the problem of evaluating different types of actions. Some have opted to delve into specific actions, while others have taken a more generalist approach. Typically, behind the metrics generated by these frameworks are objective Machine Learning (ML) tasks that predict probabilities or expected values of what is trying to be measured. [1], for example, defines the value of an action as the difference in the change of probability of scoring and conceding within the next few actions.

As explained in [17], we can benefit from exploring variants of models focused on the same task that use different levels of information. For example, by comparing the probability estimates of Expected Goals (xG) models trained with and without the goalkeeper's positioning, the difference in prediction can be attributed to how the model estimates the impact of such attributes on the probability of scoring.

[2] also approaches the problem of valuing pressing actions similarly. Their pressing framework is based on attributing to pressing the difference in probability estimates between models that use and do not use pressing features.

Likewise, we will do the same to estimate the impact of the defensive off-ball dynamics on the probability of recovering the ball. Let a game state $S_i$ be the sequence of all past actions $(a_i, ..., a_0)$ from the current action. Let $(S_i, A)$ be a game state described by feature-set $A$, which only uses action-derived features, and $(S_i, A \cup T)$ a game state described by feature-set $(A \cup T)$, that uses the same action-derived features plus tracking-derived features.

Let $P_{recovery}(S_i, F)$ be the probability of a ball recovery happening in the near future of game state $S_i$, described feature-set $F$. For simplification purposes, we will refer to $P_{recovery}(S_i, F)$ as $P(S_i, F)$. We will try to estimate the Defensive Dynamics Impact (DDI) of an action by calculating the following:

$$DDI(S_i) = P(S_i, A \cup T) - P(S_i, A)$$

In [2], VPEP's reward metric is defined as $\Delta P_{recovery}(S_i) = P_{recovery}(S_i, p_i) - P_{recovery}(S_i, \emptyset)$. The logic behind the formulation of DDI was inspired by it, with their main difference lying in the specificity of the former to pressing actions while DDI considers all actions.

### 2.1 Probabilistic Classifiers

This subsection describes the construction of the two probabilistic classifiers, $P(S_i, A)$ and $P(S_i, A \cup T)$. The tasks can be defined as:

**Given:** game state $S_i = [a_i, ... a_0]$, where $a_i$ is the current action.
**Estimates:** the probability $P(S_i, A)$ that the defending team will recover the ball in the near future of game state $S_i$, represented by the feature set $A$, that contains only action-derived features.

**Given:** game state $S_i = [a_i, ... a_0]$, where $a_i$ is the current action.
**Estimates:** the probability $P(S_i, A \cup T)$ that the defending team will recover the ball in the near future of game state $S_i$, represented by the feature set $A \cup T$, that contains tracking-derived features in addition to the previously used action-derived features.

Both $P(S_i, A)$ and $P(S_i, A \cup T)$ are binary classification problems. Thus, we train two probabilistic classifiers to estimate the probabilities of a ball recovery happening within the next couple of actions. We convert the actions into feature and label-vectors to use as input for our classifiers. In sections 2.2 and 2.3 we explain how we construct the features and labels, respectively. Later we will discuss our choice of probabilistic classifier.

## 2.2 Labels

Both $P(S_i, A)$ and $P(S_i, A \cup T)$ labels are constructed in the same way. A positive label (=1) is assigned if there is a ball recovery within the following $k$ actions and a negative label (=0) otherwise.

## 2.3 Features

The feature construction process is done similarly as in [1], [2], and [3]. Every feature instance represents a game state, described by different attributes depending on the classifier. The $P(S_i, A)$ classifier uses only action-derived features while $P(S_i, A \cup T)$ also uses tracking-derived features. Thus, a feature instance $x_i^A = [a_i^A, ..., a_{i-(\tau_1-1)}^A]$ is a vector of action-derived attributes from the current and previous $\tau_1 - 1$ actions, where $a_i^A$ represents the action-derived features of action $a_i$. Correspondingly, a feature instance $x_i^{A \cup T} = [a_i^A, ..., a_{i-(\tau_1-1)}^A, a_i^T, ..., a_{i-(\tau_2-1)}^T]$ is a vector of action-derived attributes from the current and previous $\tau_1 - 1$ actions and of tracking-derived features from the current and previous $\tau_2 - 1$ actions, where $a_i^A$ represents the action-derived features of action $a_i$ and $a_i^T$ represents the tracking features of action $a_i$. The distinction between $\tau_1$ and $\tau_2$ is because the game state representation by each of the feature-sets does not need to be the same. We will later discuss our choices.

2.2.1 Action Features

The features derived from actions are composed of the SPADL, complex, and game-context features presented in [2]. SPADL features are the raw attributes from the SPADL representation, except for the identifications, which include the action's starting and ending positions, the action type, its result, and the body part the player used to perform it. Unlike in [2], we opted not to consider the action type and result of the current action. The reason is that by including such attributes, information about the outcome may end up being encoded in the feature-vector. For example, a carry necessarily has a successful result and is a consequence of the player's decision about what to do with the ball. Correspondingly, a missed pass necessarily causes a ball recovery by the defensive team. The complex features comprise the distance and angle to the goal from the action's starting and ending positions and the total length of the action on both the x and y-axis. Finally, the game context features include the scoreline for the team in possession and out of possession and the goal difference after action $a_i$.

2.2.2 Tracking Features

From the 360º data, we construct Pitch Control (PC)[14] features to try to capture some of the off-ball and spatial control context of what is happening on the pitch. From [3], we used the average PC in a 4-meter radius around the ball. Also, we created features to try to capture the spatial occupation of the $n_{att}$ attacking and $n_{def}$ defending players with the most Relevant Pitch Control (RPC)[14] in the frame. For each of the $n_{att}$ and $n_{def}$ players, we calculate their average PC in spaces where its value is at maximum $p$ and its area on the pitch. The area is represented by the number of cells on a $32 \times 50$ grid with PC value up to $p$. We chose a set of $p$ values of $\{0.01, 0.1, 0.25, 0.5, 0.75\}$. The ordering of players by their relevant pitch control is to counter the limitation of fixed vectors of tabular data Machine Learning algorithms. Furthermore, we included the players' x and y-coordinates and their distance to the ball carrier.

# Methodology

The work was done over 580 games of Statsbomb's event and 360º data from the English Premier League's 2020/21 and 2021/22 seasons, comprising games from ten teams during those seasons. As noted in [7], one of Football Analytics' most significant challenges is "Small Sample Sizes". Since our dataset comprises 580 games, we used the data from the first 80% of the games as the training set and the final 20% for validation. Later, we used the trained models to predict the probabilities for all 580 games.

# Experiment and Results

### 4.1 Design Choices

In this subsection, we will detail the design choices of our work, especially the user-defined parameters of the framework. First, we discuss our choice of the probabilistic classifier. As described above, the feature construction requires setting the game state size of the action features ($\tau_1$) and tracking features ($\tau_2$), and the number of attacking ($n_{att}$) and defending ($n_{def}$) players. Also, the labels' construction requires defining the action window ($k$) in which a ball recovery can be observed.

4.1.1 Probabilistic Classifier

Due to its performance and accuracy, gradient boosting decision trees have become a popular technique to solve tabular data science problems. It can learn very complex non-linear relationships in the data. XGBoost (XGB) [18] is a gradient boosting decision tree model with a library and easy-to-use API. For those reasons, we decided to use it in this work. A limitation of XGB is that due to the size and complexity of learned trees, it can be a challenge for humans to interpret the model, making it a black-box model.

4.1.2 Number of actions considered per feature-type in a game state ($\tau_1$ and $\tau_2$)

We define $\tau_1$ and $\tau_2$ as user-selected parameters that represent the number of subsequent actions described by action and tracking features in the current game state, respectively. Like [1] and [2], we set $\tau_1$ to 3. It is a trade-off between adding extra information that might be useful for the predictive task versus adding redundant information that will only make the task computationally more expensive.

Unlike $\tau_1$, the choice of $\tau_2$, combined with $n_{att}$ and $n_{def}$, affects the sample size of $P(S_i, A \cup T)$. By setting $\tau_2$ to a value larger than 1, for a game state to be considered, all $(a_i, ..., a_{i-(\tau_2-1)})$ actions must satisfy $n_{att}$ and $n_{def}$. Given our data volume constraint, we opted for $\tau_2$ to be equal to 1. Thus, only the current action would need to satisfy $n_{att}$ and $n_{def}$.

4.1.3 Number of players per team in an action for tracking features

As mentioned earlier, to create fixed-length tracking features, we must choose the number of attacking and defending players for which we will build the attributes. In turn, it creates a minimum amount of attacking and defending players a frame must possess to be considered. To avoid situations where the spatial representation of the game context is insufficient and unbalanced between teams, we chose $n_{att}$ and $n_{def}$ to be 5.

4.1.4 Ball-recovery action-window label

The choice of $k$ for label construction defines the window of future actions in which we will observe a potential ball recovery. Varying $k$ implies trying to learn models that predict shorter or longer-term changes in possession. Like in [2], we tested trained models for varying $k$ to decide which to use. Table 1 informs the average prediction for each classifier and their average absolute difference.

Table 1. The table above shows the average prediction for each classifier and their average absolute difference depending on $k$.

| $k$ | $\mu(P(S_i, A))$ | $\mu(P(S_i, A \cup T))$ | $\mu(P(S_i, A)) - P(S_i, A \cup T))$ |
| --- | --- | --- | --- |
| 1 | 0.0531 | 0.0528 | 0.0150 |
| 2 | 0.1045 | 0.1037 | 0.0237 |
| 3 | 0.1540 | 0.1538 | 0.0304 |
| 4 | 0.2019 | 0.2017 | 0.0343 |
| 5 | 0.2474 | 0.2478 | 0.0376 |
| 6 | 0.2903 | 0.2907 | 0.0393 |
| 7 | 0.3301 | 0.3307 | 0.0410 |
| 8 | 0.3673 | 0.3680 | 0.0415 |
| 9 | 0.4021 | 0.4028 | 0.0420 |
| 10 | 0.4348 | 0.4355 | 0.0416 |

Table 1 is very insightful, especially by comparing it to Table 1 in [2], given the objective of each framework. In VPEP, the ball recovery classifier that does not consider pressing features always has a lower average than the one that does. This behavior is expected, as the classifier is able to learn that pressing is typically associated with ball recoveries. In our work, since we consider every action, not only those where there is pressure on the player with the ball, the probability difference between the two classifiers is always around zero. Meaning it learns both situations where the tracking features increase the probability of a ball recovery and the opposite, where the tracking attributes make the model predict a lower probability than the base classifier. This generalized behavior that would not only work in pressing scenarios was one of our main objectives when proposing the project.

As we want to predict a shorter-term change in possession, as it is what we believe the collective off-ball dynamics a team performs aims at doing, forcing a ball-recovery within a short amount of actions. Our choice of $k$ also needs to be one with a fair difference between the classifiers. Combining that with the fact that [2] chose $k = 4$ for their ball-recovery model, we decided to stick with the choice.

**4.2 Predictions' evaluation**

At the ground level of the framework are the two probabilistic classifiers. The three most used evaluation metrics for probabilistic classifiers are the Area Under the Receiver Operator Curve (AUROC), the Brier Score (BS), and Logarithmic Loss (LL). Both the BS and LL become more interpretable when normalizing, dividing their value by the BS or LL value of a baseline that predicts the class distribution. Thus, the Normalized Brier Score (NBS) and Normalized Logarithmic Loss (NLL) are generated. [20] provides an excellent discussion of which metric best suits each scenario. They recommend using the BS or NBS when the application involves summing or subtracting probabilities. As our metric is obtained by subtracting the probabilities estimated by the classifiers and other applications require adding those values, we will use the NBS as our performance evaluation metric.

Table 2. The table above reports the NBS for the models trained on different $k$ values. As we can see, they have very similar predictive performance.

| $k$ | $NBS_A$ | $NBS_{A \cup T}$ |
|---|---|---|
| 1 | 0.71 | 0.71 |
| 2 | 0.82 | 0.81 |
| 3 | 0.85 | 0.85 |
| 4 | 0.87 | 0.87 |
| 5 | 0.89 | 0.88 |
| 6 | 0.90 | 0.89 |
| 7 | 0.91 | 0.91 |

**4.3 Intuition Behind the Predictions**

The following figures exemplify how $P(S_i, A)$ and $P(S_i, A \cup T)$ work in a sequence of actions. The figures are from the Liverpool vs. Leeds United match, which took place on the 12th of September of 2020 at Anfield. The first action is made by Virgil van Dijk, a Liverpool defender, who makes a wrong pass. Kalvin Phillips, a Leeds United midfielder, has the ball after van Dijk's mistake. The third action is a pass made by Hélder Costa, also from Leeds. In those first three actions, we can see that $P(S_i, A)$ does not vary much.

However, a ball recovery by Leeds United occurs from the first to the second action. It is possible to see that the fourth action, Kalvin Phillips' pass, increases the $P(S_i, A)$, indicating a much riskier action (it is the longest and more centralized pass of the set). The last action shows Jordan Henderson, a Liverpool midfielder, carrying the ball, which indicates that a ball recovery was made.

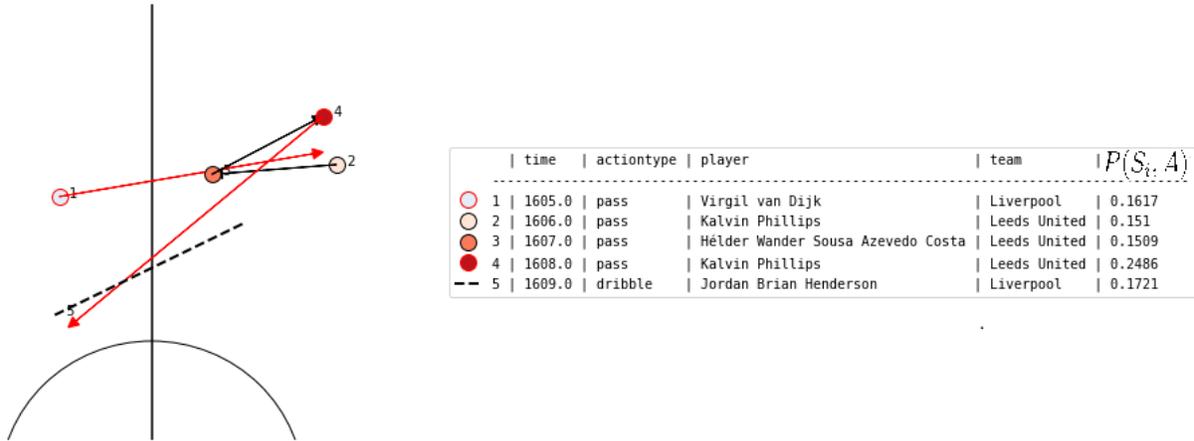

Figure 1. Liverpool vs. Leeds United (12th of September of 2020, Anfield) sequence of actions.

For the same actions, we now take into account the $P(S_i, A \cup T)$ behavior. We plot the same sequence of actions as before. However, we display it with the frame's Pitch Control. The team with the ball is indicated in red, whereas the team in blue is the one that does not have possession. In order to simplify understanding, the team with the ball will always be attacking to the right side. Therefore, Leeds' actions will be inverted compared to the previous figure. Areas in white close to players represent areas where the probability of control between teams is about even. Areas in white outside the region of play are outside the visible area reported by the data. The ball is represented as a black circle.

In the first action, the player on the ball, Virgil van Dijk, is pressed by a Leeds player. $P(S_i, A \cup T)$ identifies that, being higher than $P(S_i, A)$. The same happens in the second action, as Kalvin Phillips is making a pass pressed by a Liverpool player, in a zone of ball dispute (Pitch Control of approximately 0.5). In the next moment, Hélder Costa is less pressed and can make a clear pass, as captured by $P(S_i, A \cup T)$, that gets lower than $P(S_i, A)$. When Kalvin Phillips has the ball back, we can see that he has a bit less space than his partner, but has just enough, which makes $P(S_i, A \cup T)$ increase, yet not more than $P(S_i, A)$. As he makes a pass mistake, Henderson makes a run exactly where Liverpool have plenty of control, which lowers $P(S_i, A \cup T)$.

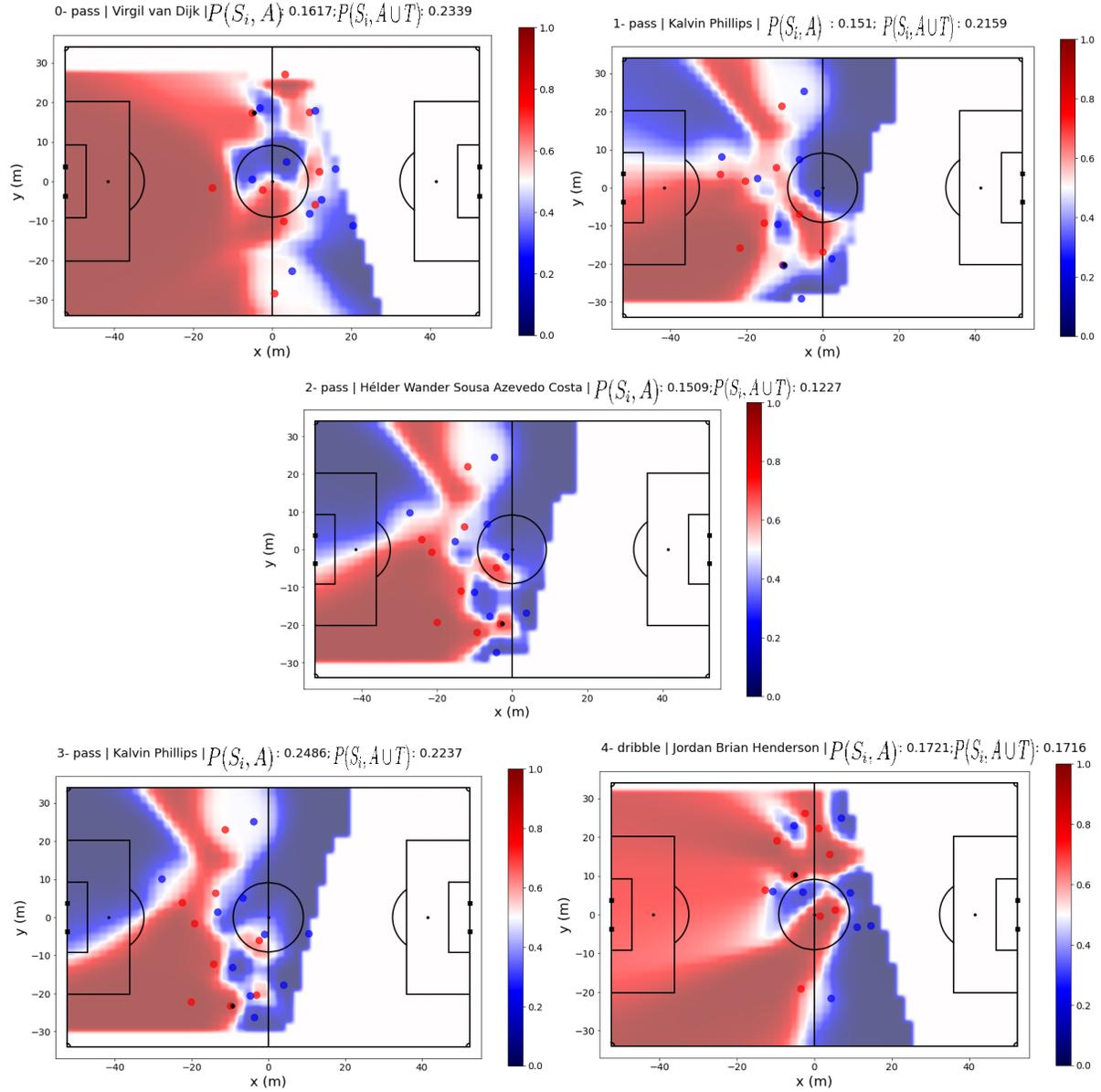

Figure 2. Liverpool vs. Leeds United (12th of September of 2020, Anfield) pitch control plot.

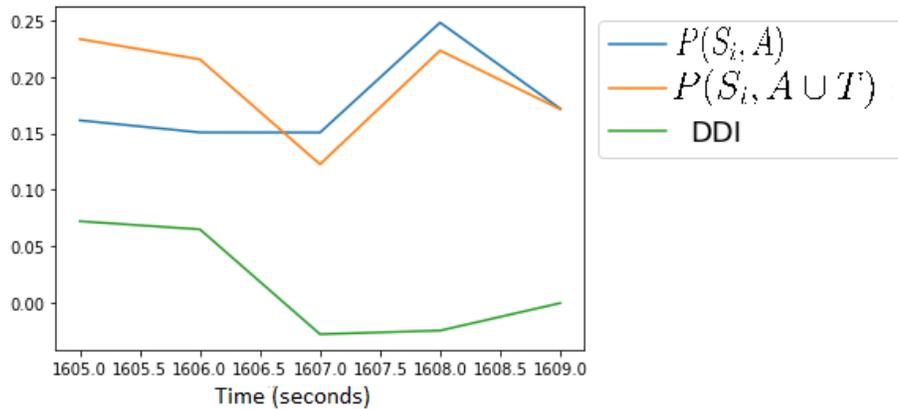

Figure 3. Liverpool vs. Leeds United (12th of September of 2020, Anfield) line graph of $P(S_i, A)$, $P(S_i, A \cup T)$, and $DDI(S_i)$ for a set of actions.

## Applications

This section will explore some possible applications of the GABR framework. An analyst can extract many possible applications from $P(S_i, A)$, $P(S_i, A \cup T)$, and DDI, both on a player or team level. We will first evaluate teams based on their accumulated DDI and characterize the pitch areas where their positioning helps recover possession. Later, we will focus on ball carriers, providing insight into the resistance, or lack of, at overcoming situations in which models predict it is likely to turn the ball over.

**5.1 Teams' Insights**

With the metrics we have constructed, there are plenty of analyses on a team's performances that we can now measure, mainly on the defensive side of the game. By calculating the mean DDI of every team's actions, we can capture which clubs usually have their probability of recovering the ball increased by their positioning.

Table 3. The mean DDI value for the top 5 teams.

| Position | Team | Mean DDI |
| --- | --- | --- |
| 1 | Leeds United | 0.007540 |
| 2 | Liverpool | 0.006952 |
| 3 | Manchester City | 0.006837 |
| 4 | Chelsea | 0.005504 |
| 5 | Southampton | 0.002526 |

From Table 3, we can see that teams usually associated with aggressive defensive styles and high pressure, such as Liverpool, Manchester City, and Leeds United, have the highest mean DDI, indicating that the model can capture pressing moments.

Another interesting insight is to see where on the pitch clubs usually have a higher or lower DDI. With such information, managers can prepare tactics based on their opposition's defending style. In the Figure below, it is possible to see how Leeds United over performs Newcastle United, the team with the least mean DDI, on almost every part of the pitch, with a much higher DDI, especially in the central sector on the offensive half (attacking play to the right).

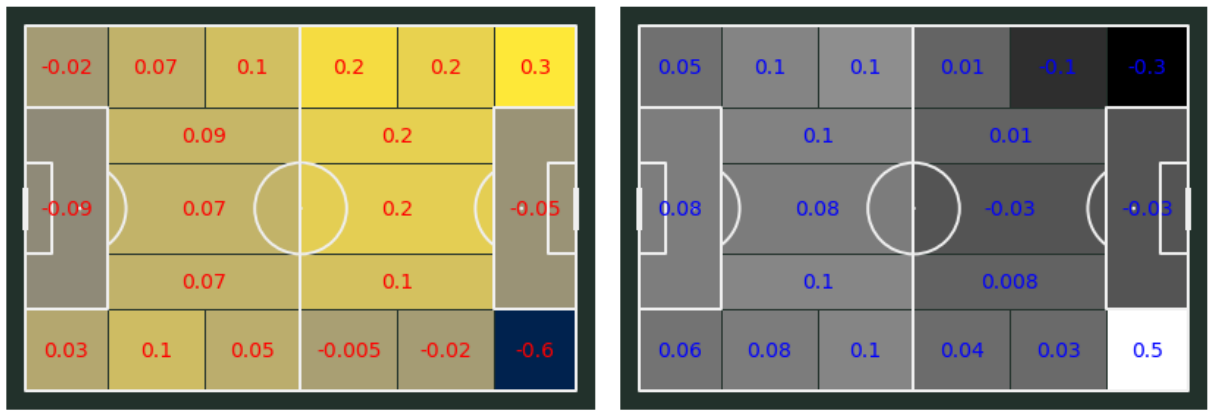

Figure 4. Leeds United (left) and Newcastle United (right) mean DDI on positional pitch.

As the methodology section describes, our data comprises 2020/21 and 2021/22 English Premier League games, in which Leeds United's manager was Marcelo Bielsa. As 'El Loco' (Bielsa's nickname) was sacked before the end of the 2021/22 season, we can compare his knowingly aggressive pressure style with his successor in the job, Jesse Marsch, and see if DDI captures Bielsa's team characteristic.

Table 4. Bielsa vs. Marsch - Leeds United DDI.

| Manager | Club | Period | # of Games | Mean DDI |
|---|---|---|---|---|
| Marcelo 'El Loco' Bielsa | Leeds United | 01/07/2021 - 27/02/2022 | 64 | 0.008810096 |
| Jesse Marsch | Leeds United | 27/02/2022 - 22/05/2022 | 12 | 0.0022783908 |

### 5.2 Players' Insights

As Statsbomb 360° Data does not identify the players other than the ball carrier, we can not analyze a player's defensive or offensive off-ball performance. However, we analyze how players with the ball perform under scenarios of more or less ball-recovery propensity.

For example, players that consistently perform actions in which $P(S_i, A \cup T)$ is above a certain threshold (0.90, for example) and the associated label is negative are probably excellent possession retainers. Table 5 displays the top five players with the most number of actions under this condition.

As we can see in Table 5, some of the best possession retainers are present. The exception to such a trait would be Raphinha, which is likely on the list due to surpassing defenders in wing situations where he is very likely to lose the ball.

Table 5. Displays the players with the most actions satisfying: $P(S_i, A \cup T) > 0.9$ and a negative label.

| Position | Player | Team | Count |
|---|---|---|---|
| 1 | Bruno Fernandes | Manchester United | 96 |
| 2 | Kevin De Bruyne | Manchester City | 80 |
| 3 | Harry Kane | Tottenham Hotspur | 80 |
| 4 | João Cancelo | Manchester City | 73 |
| 5 | Raphinha | Leeds United | 70 |

On the contrary, players that consistently perform actions in which $P(S_i, A \cup T)$ is below a certain threshold (0.10, for example) and the associated label is positive are probably players that turn the ball over a lot, be it because of skill or style. For example, Trent Alexander-Arnold is a player known for consistently trying to verticalize the game. Since lots of these attempts at finding a longer pass originate from low possession conceding probabilities, he would definitely rack up these types of situations.

Another possible use of our framework is to identify 'Hospital Balls' [16] and players that do it the most. The identification of a hospital ball can be by simply looking at situations where the $P(S_i, A \cup T)$ difference between the current and the following game state is significant, and there is a possession change. Table 6 displays the players that do it most.

Table 6. Top 5 players that performed most actions where $P(S_{i+1}, A \cup T) - P(S_i, A) \geq 0.75$ (count per 100 actions for those who have more than 10 actions of this nature and where $a_{i+1}$ is performed by a teammate).

| Position | Player | Team | Count | Count/100 actions |
|---|---|---|---|---|
| 1 | Bruno Fernandes | Manchester United | 19 | 0.227572 |
| 2 | Harry Maguire | Manchester United | 14 | 0.202312 |
| 3 | João Cancelo | Manchester City | 20 | 0.189915 |
| 4 | Ashley Westwood | Burnley | 11 | 0.187011 |
| 5 | Yves Bissouma | Brighton & Hove Albion | 11 | 0.179944 |

## Conclusion and Future Work

The Generalized Action-based Ball Recovery framework allows insight into the defensive dynamics a team deploys to recover the ball. To construct the framework, we built two generalized ball recovery classifiers that account for different levels of information, $P(S_i, A)$ and $P(S_i, A \cup T)$, which by themselves have proven to also be useful. It is a generalization of the work done in [2] in the sense that it considers all actions, not only pressing scenarios. We went further and constructed a novel metric based, DDI, based on the two classifiers, which gives a perception of how well a team is positioned on the pitch. The applications of the paper range from opponent analysis to scouting at team and player levels. We proposed a set of Pitch Control features that seek to capture the spatial occupation of the players most likely to receive a pass and fit well with tabular data Machine Learning algorithms.

We plan on exploring new Pitch Control-based features to see if they can improve the prediction performance of the $P(S_i, A \cup T)$ classifier, as we believe it has that potential. Applying both $P(S_i, A)$ and $P(S_i, A \cup T)$ as parts of other frameworks would be another possible direction. Finally, as another contribution of the work, we plan on releasing a Cython version of the 'Friends of Tracking' Pitch Control [13][14] implementation with its adaptations to Statsbomb 360º Data. We believe not many public works (not necessarily papers) have explored Pitch Control-based analysis on 360º Data due to the troubles of adapting and scaling the available implementation.